\documentclass[a4paper]{article}

\usepackage{INTERSPEECH2022}

\usepackage{enumitem}
\usepackage{slantsc}
\usepackage{multirow}
\usepackage{dsfont}
\usepackage{color}
\usepackage{arydshln}
\usepackage{booktabs}
\usepackage[export]{adjustbox}
\usepackage[lofdepth,lotdepth]{subfig}
\usepackage[tracking=true]{microtype}
\usepackage{cite}
\usepackage{xcolor}
\usepackage{hyperref}
\hypersetup{colorlinks,allcolors=black}
\usepackage{xspace}
\usepackage{stfloats}

\title{\vspace{-1.5ex}Speech Quality Assessment through MOS using Non-Matching References\vspace{-0.7ex}}
\name{Pranay Manocha$^{1*}$\thanks{$^{*}$Work done during internship at Meta.},
      Anurag Kumar$^{2}$}
\address{
  $^1$Department of Computer Science, Princeton University, Princeton, NJ, USA\\
  $^2$Meta Reality Labs Research, Redmond, WA, USA}
  
\email{{pmanocha@cs.princeton.edu}, {anuragkr90@fb.com}\vspace{-1ex}}
\DeclareUnicodeCharacter{2212}{-}
\newcommand{\ignorethis } [1] {}

\newcommand{\etal       }     {{et~al.}}

\newcommand{\eg         }     {{e.g.}}

\newcommand{\ie         }     {{i.e.}}

\newcommand{\Reals      }     {{\textrm{I\kern-0.18em R}}}

\renewcommand{\vec      } [1] {{\text{\boldmath $\mathbit{#1}$}}}

\newcommand{\change     } [1] {\mbox{{\footnotesize $\Delta$} \kern-3pt}#1}

\newlength{\w}

\newcommand{\acronym}[1] {{\uppercase{#1}}}

\newcommand{\PESQ}   {\acronym{Pesq}}
\newcommand{\VISQOL} {\acronym{Visqol}}
\newcommand{\POLQA}  {\acronym{Polqa}}

\newcommand{\NISQA}   {\acronym{Nisqa}}

\newcommand{\OURS}   {\acronym{NORESQA-MOS}}

\newcommand{\NORESQA}   {\acronym{Noresqa}}

\newcommand{\VOICEMOS}  {\acronym{Voicemos}}
\newcommand{\BVCC}  {\acronym{Bvcc}}
\newcommand{\DNSMOS}  {\acronym{Dnsmos}}
\newcommand{\DMOS}  {\acronym{D-MOS}}

\makeatletter
\newcommand*{\etc}{%
    \@ifnextchar{.}%
        {etc}%
        {etc.\@\xspace}%
}
\makeatother

\begin{document}

\maketitle

\begin{abstract}
Human judgments obtained through Mean Opinion Scores (MOS) are the most reliable way to assess the quality of speech signals. However, several recent attempts to automatically estimate MOS using deep learning approaches lack robustness and generalization capabilities, limiting their use in real-world applications. In this work, we present a novel framework, \OURS, for estimating the MOS of a speech signal. Unlike prior works, our approach uses non-matching references as a form of conditioning to ground the MOS estimation by neural networks. We show that \OURS\ provides better generalization and more robust MOS estimation than previous state-of-the-art methods such as \DNSMOS~\cite{reddy2020dnsmos} and \NISQA~\cite{mittag2021nisqa}, even though we use a smaller training set. Moreover, we also show that our generic framework can be combined with other learning methods such as self-supervised learning and can further supplement the benefits from these methods. 
\end{abstract}

%Many existing full-reference models correlate well with human perception, but cannot be used in real-world scenarios where ground truth clean reference recordings are not available. On the other hand no-reference metrics typically suffer from several shortcomings, such as lack of robustness to unseen perturbations and reliance on (limited) labeled data for training. The \NORESQA\ approach of Manocha et al.~\cite{manocha2021noresqa} learns a non-matching reference metric trained directly on a multi-task network of predicting SNR and Si-SDR directly. However, these pseudo metrics may not directly correlate well with human perception. This paper introduces \OURS\ - a metric that builds on top of the NORESQA framework by estimating relative MOS ratings. The primary improvement is to directly train on relative human judgments of absolute quality (MOS) to build robust models using limited labeled data. We show that neural networks trained in this manner generalize much better across different OOD tasks.

% computational models for human auditory perception. 

\noindent\textbf{Index Terms}: speech quality, non-matching reference, Mean Opinion Score, no-reference metrics, speech enhancement

\section{Introduction}

% computational models for human auditory perception. 
Quality assessment of speech signals plays a critical role in many applications. The gold standard for assessment of speech quality is subjective judgments by humans. Often, these subjective judgments are made by conducting different listening tests. Mean Opinion Score (MOS)~\cite{streijl2016mean} is the ``de-facto'' metric to assess speech quality through listening tests. However, such subjective evaluations are time and resource consuming, especially when repeated many times per recording, and are therefore not scalable. Moreover, to obtain MOS reliably, one needs to control listening environments and hardware appropriately, further adding to the constraints of conducting MOS tests. This has led to considerable effort in developing alternatives to MOS tests.

%recent times to develop deep learning based MOS estimation methods {\bf [CITE-ALL]}. 

%Speech quality assessment (SQA) plays a fundamental role in many applications affecting the quality of listening experiences.
%
%Gold standard assessments of speech quality - Mean Opinion Score (or MOS) requires subjective listening tests. 
%

% notes for references
% limitationas objective measures only serve as approximations of human assessment [15, 16] -- cite 15 and 16 from https://zzhang68.github.io/papers/Zhang.Assessment.ICASSP2021.pdf
% These measures,however, do not always correlate well with subjective quality results [12, 13] -- https://arxiv.org/pdf/2007.15797.pdf

One class of alternatives that have been developed are full-reference objective methods, \eg\ \PESQ\cite{rix2001perceptual}, \POLQA\cite{beerends2013perceptual} and \VISQOL\cite{hines2015visqol}), to mention a few. While these objective metrics remove the heavy workload of subjective listening tests, they correlate with MOS to a limited degree~\cite{cano2016evaluation, emiya2011subjective, cooper2021generalization}.
More importantly, their effectiveness is usually limited to specific speech applications and becomes obsolete with the emergence of new scenarios~\cite{hines2013robustness,manjunath2009limitations}. Even more inhibiting is the reliance of these objective metrics on a clean, reference speech signal for computing an assessment rating.

A recent class of alternatives is provided by deep-learning-based systems, which offer scalable and rapidly re-trainable solutions that are expandable to many speech and audio-related tasks~\cite{manocha2020differentiable,manocha2021cdpam,lo2019mosnet,patton2016automos,fu2018quality,fu2019metricgan,yu2021metricnet}. Several of these methods estimate the aforementioned objective metrics (\eg\ \PESQ~\cite{rix2001perceptual}) directly, without using any reference. More significantly, there have also been attempts to learn the mapping between audio signals and MOS directly.

%Most of these methods rely purely on supervised learning~\cite{reddy2020dnsmos,zhang2021end,catellier2020wawenets,mittag2021nisqa}. However, these methods often fail to generalize well to unseen conditions 

The task of developing machine learning methods for MOS estimation is quite challenging. MOS captures the complex and multi-dimensional nature of quality perception in humans~\cite{coren2004sensation}. However, several aspects of human auditory perception are not yet fully understood. This makes it tricky to design MOS estimation methods, and often the idea is to rely on labeled MOS datasets for training neural networks in a supervised manner~\cite{lo2019mosnet,patton2016automos,mittag2021nisqa,cooper2021generalization,huang2021ldnet,reddy2020dnsmos,catellier2020wawenets,zhang2021end}. However, collecting large scale MOS datasets to train deep learning models is challenging too. Current MOS datasets are often limited to specific domains, \eg\ Text-to-Speech (TTS) and Voice Conversion in \BVCC~\cite{voicemos}, telephony distortions in \NISQA~\cite{mittag2021nisqa}, and speech enhancement distortions in \DNSMOS~\cite{reddy2020dnsmos}. Moreover, MOS tests are difficult to conduct and crowd-sourced MOS can have considerable label noise~\cite{reddy2020dnsmos}. These limitations make it harder to train models that can generalize well across various test conditions and applications~\cite{dong2019classification, manocha2021noresqa,cooper2021generalization}, and the real-world uses of these MOS estimation methods remain limited. 

A potential solution to above constraints can be self-supervised learning (SSL). SSL methods leverage large unlabeled data for learning models that can be utilized in other tasks with sparse labeled data. Cooper \etal~\cite{cooper2021generalization} proposed the same for MOS estimation by using large pretrained audio models learned using SSL methods (\eg\ wav2vec2.0~\cite{baevski2020wav2vec} and HuBERT~\cite{hsu2021hubert}). 

% SSL for audio and speech as well as MOS. 
% Introduce NORESQA. Motivation. (1) humans can compare (2) better grounding by conditioning with non-matching references. Developed for SNR and SI-SDR. 
% extend to MOS. can be combined with SSL. added benefits. 

Another recent novel framework for quality assessment is \NORESQA~\cite{manocha2021noresqa} (\textit{NO}n-matching \textit{RE}ference based \textit{S}peech
\textit{Q}uality \textit{A}ssessment). Motivated by human's ability to compare the quality of two speech signals of different content, \NORESQA\ proposed speech quality assessment by learning to predict a relative quality score for a given speech recording with respect to \emph{any} provided reference, irrespective of the differences in content, speaker's language or gender. The non-matching references (NMRs) in \NORESQA\ provide better grounding for the neural networks through conditioning by arbitrary speech signals of known quality. However, \NORESQA\ was trained to predict Signal to Noise Ratio (SNR) and Scale invariant signal to distortion ratio (Si-SDR) for quality assessment. 

In this paper, we propose \OURS\ - a novel MOS estimation method built on the principles of \NORESQA. Unlike prior works which are entirely reference-free, \OURS\ relies on random NMRs of known qualities/MOS (either from a labeled dataset, or a clean set). We show that using our approach to compute relative MOS ratings leads to high generalization across in-domain and out-of-domain datasets. %Moreover, this generic framework can be combined with other useful approaches, such as SSL pre-training. 
Moreover, combining \OURS\ with other useful approaches (\eg\ SSL pretraining) provides computational benefits by enabling smaller models to achieve significantly better generalization for MOS prediction.
%Moreover, this generic framework can be combined with other useful approaches, such as SSL pre-training.
%
%\OURS\ combined with SSL-pretraining provides computational benefits by enabling smaller models to achieve significantly better generalization for MOS prediction.
%
\OURS\ is usable in real-world applications as any other reference-free approach as one can choose any set of speech recordings as NMR inputs to the network.

\section{The NORESQA-MOS Framework}

 Our framework,~\OURS\ is designed to assess the MOS of a given speech recording using Non-Matching References (NMRs). The model takes in two recordings as inputs: a test recording $x_\mathtt{test}$ and another randomly chosen recording $x_\mathtt{ref}$. Fig~\ref{model_framework} is a simple illustration of the model. Overall, given two input signals, our model predicts two outputs: i) a preference output suggesting which input is cleaner than the other, and ii)~a relative MOS rating between the two inputs.

\subsection{Framework Design and Model Architectures}
\label{models}
\begin{figure}[t!]
\vspace{-1\baselineskip}
\centering
\setlength{\tabcolsep}{0pt}
\includegraphics[width=\columnwidth]{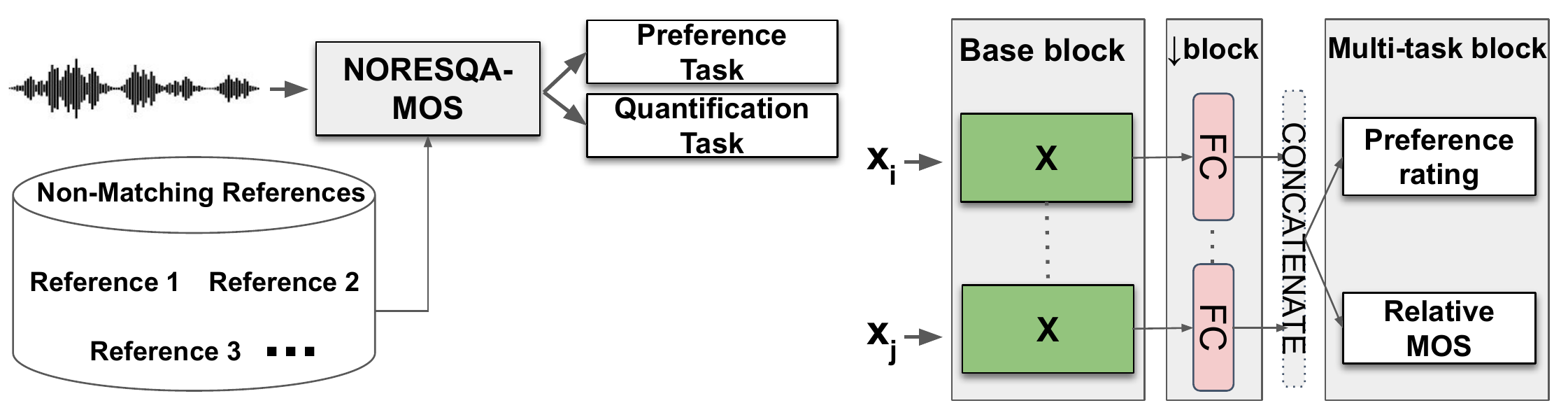}
\vspace{-4ex}
\caption{Left: \emph{\OURS\ Framework}: takes a test recording, and a randomly chosen NMR from a set and solves a preference and a quantification task.
Right: \emph{\OURS\ Architecture}: takes two inputs ($x_i$ and $x_j$), passes through a base model ($X$ = scratch or pretrained wav2vec2.0), and outputs: (i)~which recording is cleaner (preference-task); and (ii)~relative MOS score (quantitative-task).}
\vspace{-4ex}
% \vspace{-0.1in}
\label{model_framework}
\end{figure}

\OURS\ architecture (Fig~\ref{model_framework}) comprises three modules: \emph{a base model block}, \emph{a downsampling block}, and \emph{task specific output heads} (preference and relative MOS prediction blocks). 

\noindent {\bf Base model block:}
We consider two types of base model blocks: one where the base model is trained from scratch and another where the base model block is a pre-trained SSL model from Fairseq~\cite{ott2019fairseq}. Overall, we train 3 different models with same architectural design (based on wav2vec2.0~\cite{baevski2020wav2vec}) but varying model capacity: (i)~\emph{Scratch}: same model architecture as wav2vec2.0, but less number of blocks, and consists of roughly 120k parameters; (ii)~\emph{SSL-Small}: mid-size pretrained SSL wav2vec2.0 model (``wav2vec\_base'') consisting of roughly 91M parameters, and (iii)~\emph{SSL-Big}: large pretrained SSL wav2vec2.0 model (``wav2vec\_big'') consisting of roughly 315M parameters.

\noindent {\bf Downsampling block:}
Consists of a fully connected layer that outputs 32 dimensional representations for each time-frame. The learnable parameters across these blocks are shared between the two inputs to our model. Finally, the embeddings for both inputs are concatenated, and passed on to the next blocks. 

The next blocks consist of output heads for the training tasks, and are described below along with the training loss functions.

\subsection{Training Tasks and Loss Functions}
\label{subsec2.2}
We follow a multi-task learning framework where we train our network on two tasks simultaneously: i) a ~preference task, and ii)~quantification task using a multi-task learning (MTL)~\cite{caruana1997multitask} framework. Both output heads use attention pooling~\cite{mittag2021nisqa} to aggregate frame-level outputs to recording-level outputs. It mimics the selective auditory attention~\cite{koch2011switching} properties due to which quality cannot be estimated using simple averaging. 

\noindent {\bf Preference Task} is designed such that the network learns to model which of the two inputs is ``\emph{preferred}" by humans. It is formulated as a binary classification problem. Let $\mathbf{x}_{ij} = (x_i, x_j)$ be an \emph{ordered} pair input to the network, with $x_i$ as first input and $x_j$ as second input. Let $\text{MOS}_{x_i}$ and $\text{MOS}_{x_j}$ be the MOS ratings of $x_i$ and $x_j$ respectively. The goal is to predict the probability, $\mathbf{p}_{ij}$, of $x_i$ having better rating than $x_j$.  More formally, the label $\mathbf{y}_{ij}$ for $\mathbf{x}_{ij}$ is a 2 dimensional, one-hot vector, with $\mathbf{y}_{ij} = [1, 0]$ if $\text{MOS}_{x_i} > \text{MOS}_{x_j}$, and $\mathbf{y}_{ij} = [0, 1]$ otherwise. The loss function is:
\vspace{-0.1in}
\begin{equation}
\label{eq:pref_task}
L_P(\mathbf{x}_{ij}, \mathbf{y}_{ij}) = \sum_{k=1}^{2}{-y^k_{ij}\log(p^k_{ij})}
\vspace{-0.05in}
\end{equation}

\noindent {\bf Relative Rating Task}
is designed to quantify the quality difference (MOS) between $x_i$ and $x_j$. The goal of this task is to predict the relative MOS ratings, $\MakeUppercase{\Delta} \text{MOS}_{ij} = \mathbf{s}_{ij}  = |\text{MOS}_{x_i} - \text{MOS}_{x_j}|$. Let $\mathbf{r}_{ij}$  be the recording level relative MOS rating predicted by this output head. We then use L1 loss between $\mathbf{r}_{ij}$  and the target relative MOS $\mathbf{s}_{ij}$ to train the network:
\vspace{-0.1in}
\begin{equation}
\label{eq:pref_task}
L_Q(\mathbf{x}_{ij}, \mathbf{s}_{ij}) = \|r_{ij} - s_{ij}\|_{1}
\vspace{-0.05in}
\end{equation}

\subsection{Training procedure}
\label{training_procedure}
We assume the availability of a small labeled dataset of audio recordings, and their MOS ratings $\mathcal{D}_{\text{lab}}$. We also assume the availability of a clean speech database $\mathcal{D}_{\text{clean}}$.

The training input for the network, $\mathbf{x}_{ij}$, is created by sampling two recordings $x'_i$ and $x'_j$ (having MOS ratings $s_i$ and $s_j$ respectively) from $\mathcal{D}_{\text{lab}}$. 
Also note that $x'_i$ and $x'_j$ can also be sampled from $\mathcal{D}_{\text{clean}}$ whose rating is assumed to be the perfect MOS ($s_i$,$s_j$ = 5).
%
%We also sample a third recording ($x'_k$) from $\mathcal{D}_{\text{clean}}$ whose rating is assumed to be the perfect MOS ($s_k$ = 5). 
%
%We can randomly replace one of the recordings (from $x'_i$ or $x'_j$) with $x'_k$ to ensure that the model learns from a new set of clean recordings. \AKnote{Not sure I understand this line.}
%
Next, given $x'_i$ and $x'_j$, we apply data augmentations on the recordings including waveform inversion, audio reversal, and time stretching.
Typically, it has been found that data augmentation improves performance, especially in situations that have sparse labeled examples~\cite{shorten2019survey}. All these augmentations are chosen such that they have none to minimal effect on MOS ratings, and training with these augmentations improve performance.
% \AKnote{Should we mention that these augmentations are expected to have minimal effect on MOS label but helps improve performance given the limited data}
%
For each recording, we sample a perturbation from the list above, and apply the perturbation at a randomly selected level to get recordings $x_i$, and $x_j$ respectively.
Once we have the signals ($x_i$ and $x_j$) and their respective MOS ratings $s_i$ and $s_j$, we can train the network as described in Sec~\ref{subsec2.2}.

\subsection{Usage: MOS Prediction}
\label{usage}
Once the network is trained, we can predict the MOS of a test input $x_{\text{test}}$ with respect to any reference $x_{\text{ref}}$. As already mentioned, this reference need \textbf{not} be the matching clean reference. 
To obtain the ``absolute" quality, we select multiple clean NMRs (from $\mathcal{D}_{\text{clean}}$) with the assumption of perfect MOS ratings. We average the \emph{relative-rating block} outputs over multiple NMRs to obtain a lower variance estimate of MOS.

%The relative quality score, \OURS, of $x_{\text{test}}$ w.r.t $x_{\text{ref}}$ is obtained from the quantification head of the network. The~\frameworkname\, score gives us only the magnitude of difference between $x_{\text{test}}$ and $x_{\text{ref}}$. The output of the preference-task tells us the ``sign", whether $x_{test}$ is better or worse than $x_{ref}$. 

%Our method provides a way to compare the quality of any two given speech recordings. However, in practice, one might often be interested in measuring ``true" or ``absolute'' quality, which under our framework can be done by using clean or high-quality speech recordings as references. More specifically, the $x_{ref}$ samples could come from \emph{any} clean speech database. Moreover, to reduce variance in the estimate, we can sample multiple references and obtain an average~\OURS\, score where the NMRs are sampled from the database.

\section{Experimental Setup}

\subsection{Datasets and training}
The clean NMR set ($\mathcal{D}_{\text{clean}}$) comes from the DAPS dataset~\cite{mysore2014can}. The labeled MOS dataset ($\mathcal{D}_{\text{lab}}$) comes from \BVCC~\cite{cooper2021voices}. It combines audio recordings from past years' Blizzard Challenge for TTS and the Voice Conversion Challenge, with each recording being rated by 8 independant raters. 
Overall, it contains roughly 7000 audio recordings, and their corresponding MOS ratings. We use the pre-created training/development/test splits as provided by the \VOICEMOS\ challenge organizers.  
%
% It combines audio recordings from the blizzard challenge for TTS and the voice conversion challenge, with each recording being rated by 8 independant raters.
%\AKnote{Add a line that dataset is mainly TTS/Voice Conv ? If already mentioned somewhere in Exp/Results then its fine. But we need to emphasize the data challenge somewhere. }

The inputs to our model are 3 seconds waveform excerpts. We use the Adam optimizer with a learning rate of $10^{−4}$ with a batch size of 64. We train the network for 1000 epochs. We also use \emph{n}=100 NMRs for all evaluations.

\subsection{Baselines}
We compare our approach to state-of-the-art no-reference approaches like \DNSMOS~\cite{reddy2020dnsmos} and \NISQA~\cite{mittag2021nisqa}.
Moreover, for a fair comparison and to demonstrate effectiveness of our NMR based approach, we also compare it with a model that is exactly same as ours but predicts the absolute MOS directly (\DMOS, short for Direct-MOS). Also note that all models are evaluated at 16kHz except \NISQA\ which predicts MOS at 48kHz.

\begin{figure}[t!]
\vspace{-1\baselineskip}
\centering
\setlength{\w}{0.48\columnwidth}
\setlength{\tabcolsep}{1pt}
\begin{tabular}{cc}
\includegraphics[width=\w]{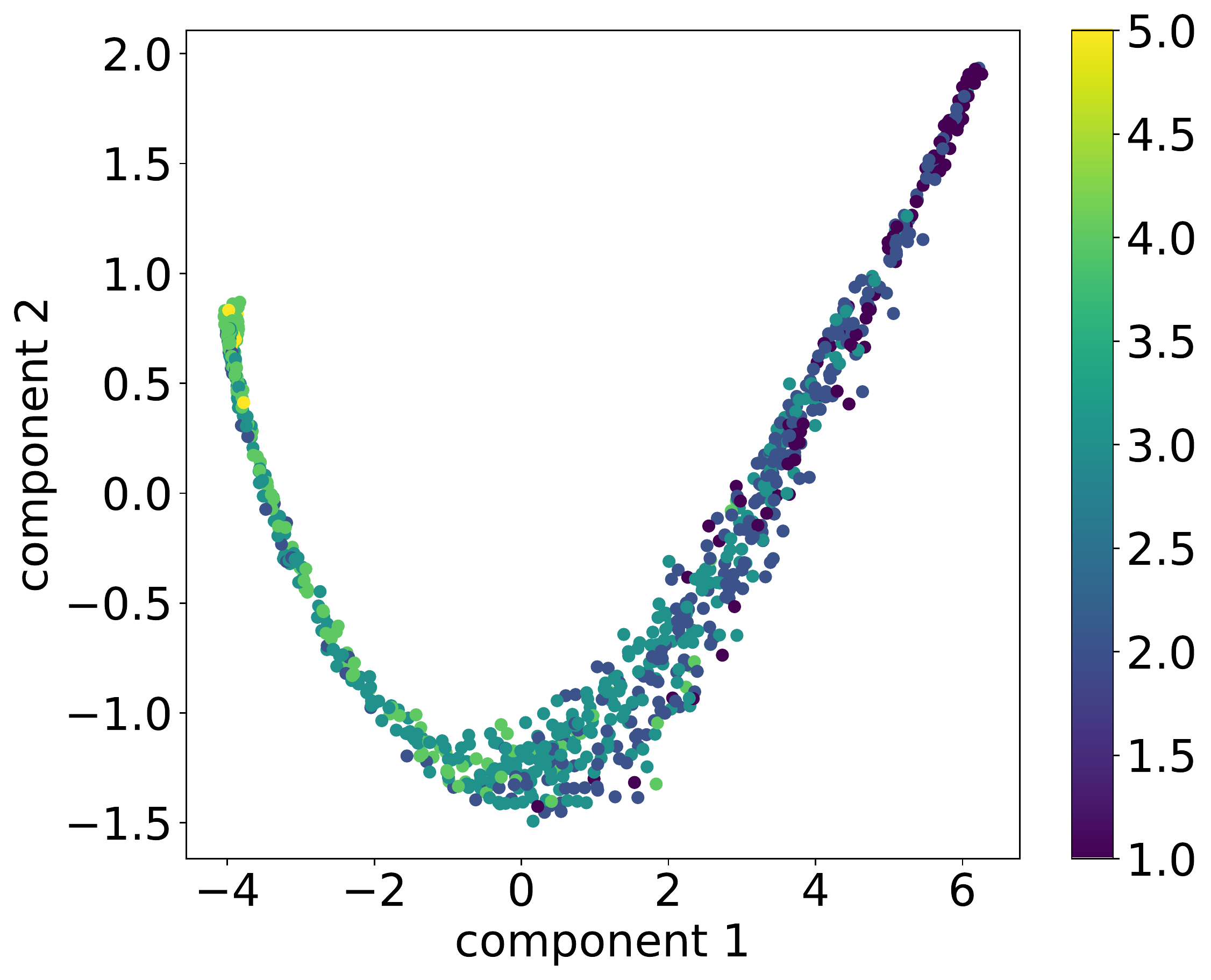} &
\includegraphics[width=\w]{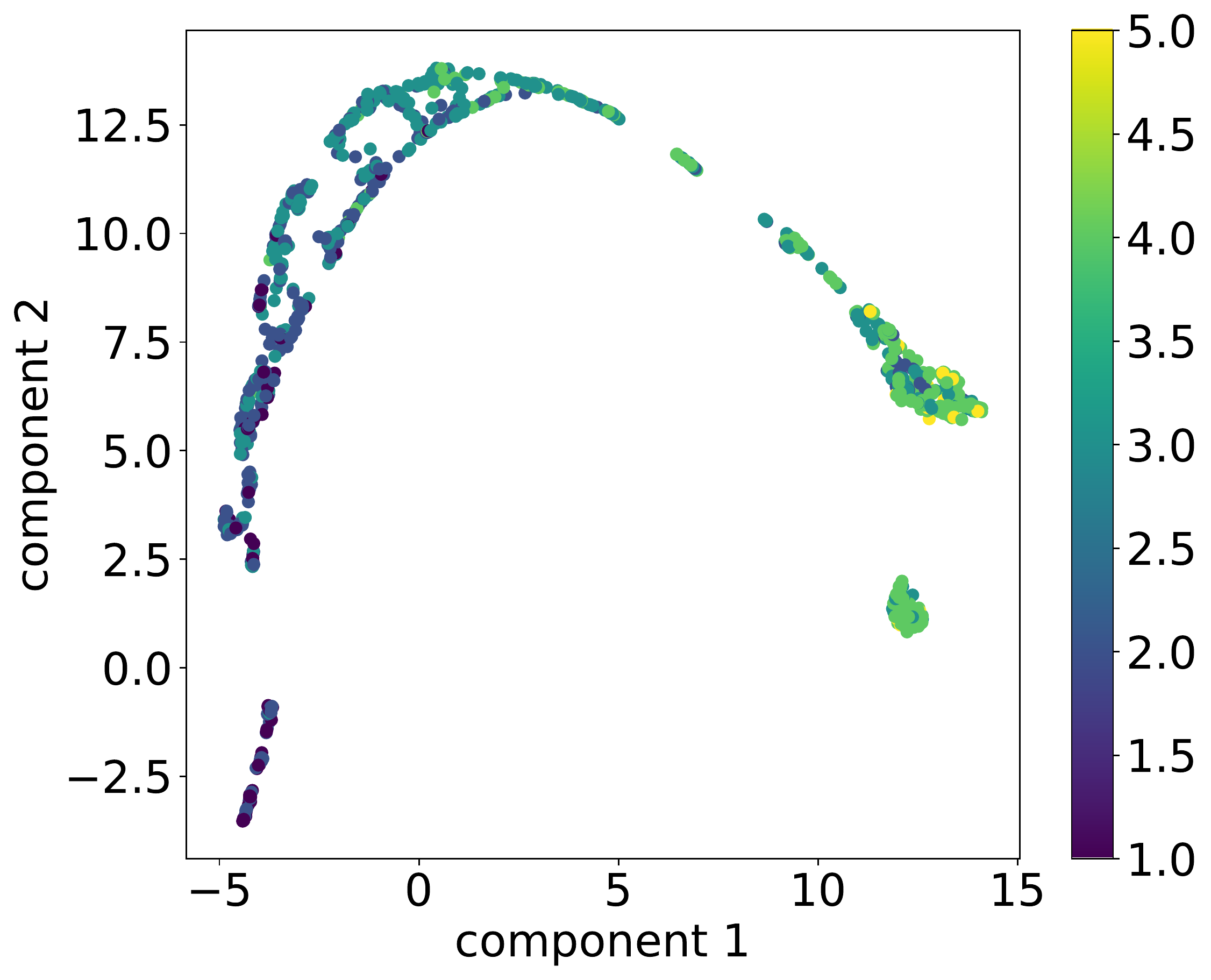} \\

\begin{minipage}{\w}\centering
\vspace{2pt}
{\footnotesize {\bf (a) PCA} }

\end{minipage} &
\begin{minipage}{\w}\centering
\centering
{\footnotesize {\bf (b) UMAP}} 
\end{minipage} 
\end{tabular}
%\vspace{-3.5ex}
\caption{\textbf{Embedding Visualization}: (a)~PCA visualization showing local structure; and (b)~UMAP visualization showing global structure. Plots show that the embeddings captures audio quality information.}
\label{embeddings_plot}
\vspace{-4ex}
\end{figure}

\section{Results}

\subsection{Objective evaluations}
\label{obj_eval}
We conduct two objective evaluations to understand the embedding space learnt by \OURS. We first look at how well the model clusters audio recordings of similar MOS ratings. Next, we visualize the embedding space of \OURS\ to see if the model learns local or global structure.

\noindent {\bf Quality based retrieval}:
Here we consider the outputs after the base model block as the quality embeddings, and use it for quality based retrievals. Similar to~\cite{manocha2021noresqa}, we first create a test dataset of 1000 recordings at 10 discrete quality levels (from 1 to 5). We take randomly selected queries and calculate the number of correct class instances in the top $K$ retrievals. We report the mean of this metric over all queries ($\text{MP}^k$). \OURS\ gets $\text{MP}^{k=10}$ = 0.92, as compared to \DMOS\ $\text{MP}^{k=10}$ = 0.85, suggesting that our approach better clusters quality level groups in this learnt space.
%these quality level groups all cluster together in this learnt space. 
%\AKnote{Any chance you can add numbers for the Abs. MOS model. Not a biggy but it will make it more complete, especially if NORESQA-MOS is better.}

\begin{figure*}[b!]
\vspace{-1.4\baselineskip}
\centering
\setlength{\w}{0.50\textwidth}
\setlength{\tabcolsep}{5pt}
\renewcommand{\arraystretch}{-1}

\begin{tabular}{cc}
\hspace{-0.3in}
\includegraphics[width=\w]{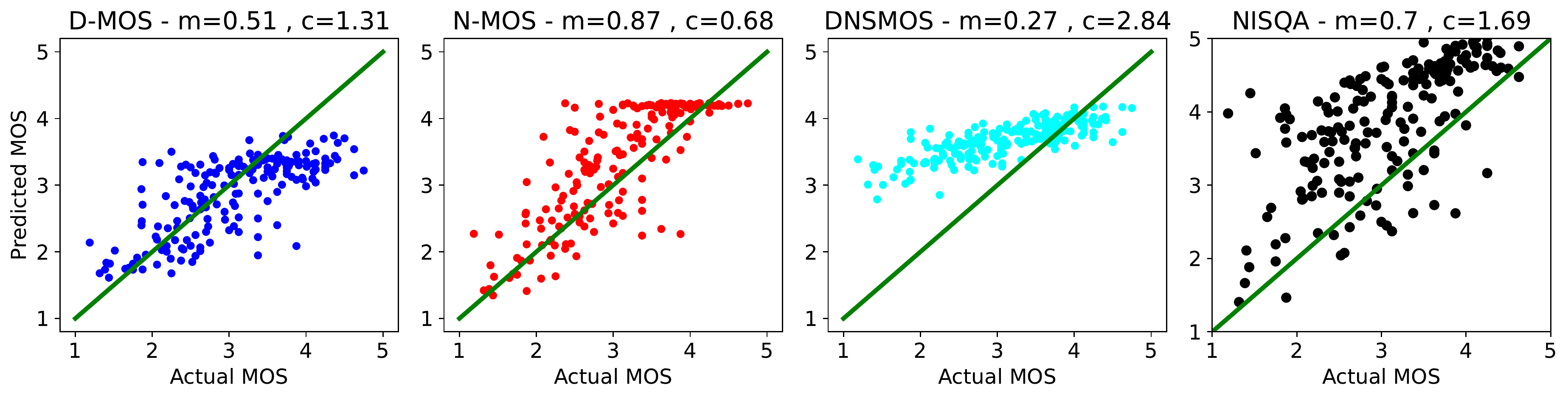}&
\hspace{-0.3in}
\includegraphics[width=\w]{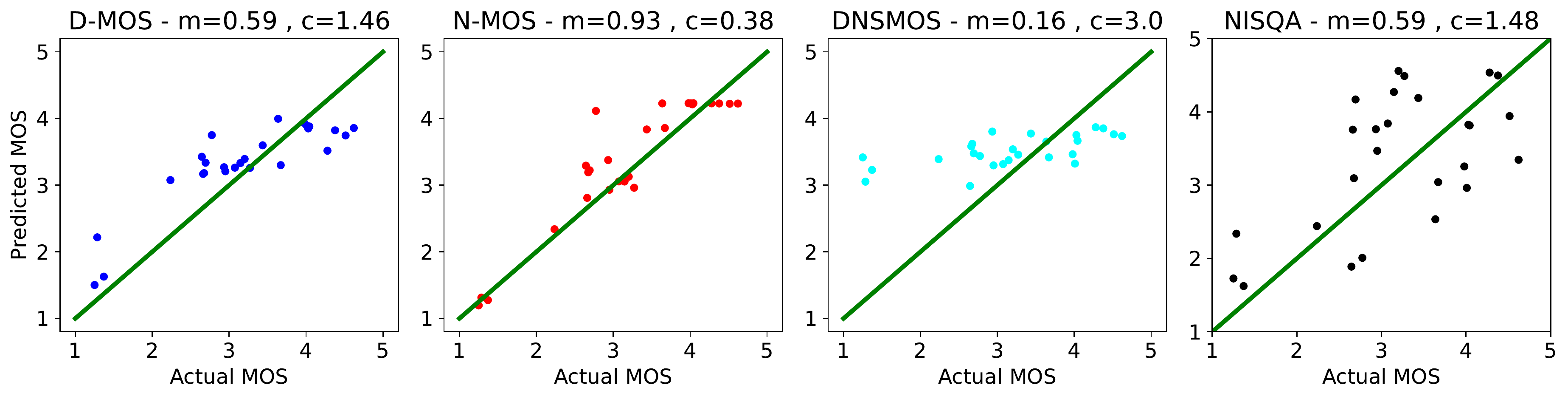} \\

\hspace{-0.3in}\includegraphics[width=\w]{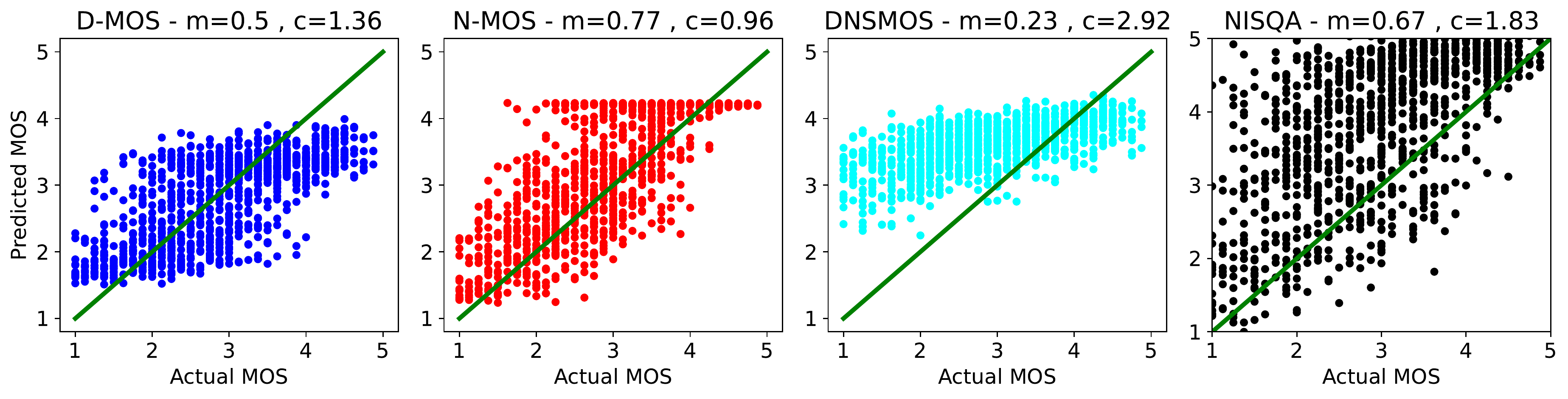} & 
\hspace{-0.3in}
\includegraphics[width=\w]{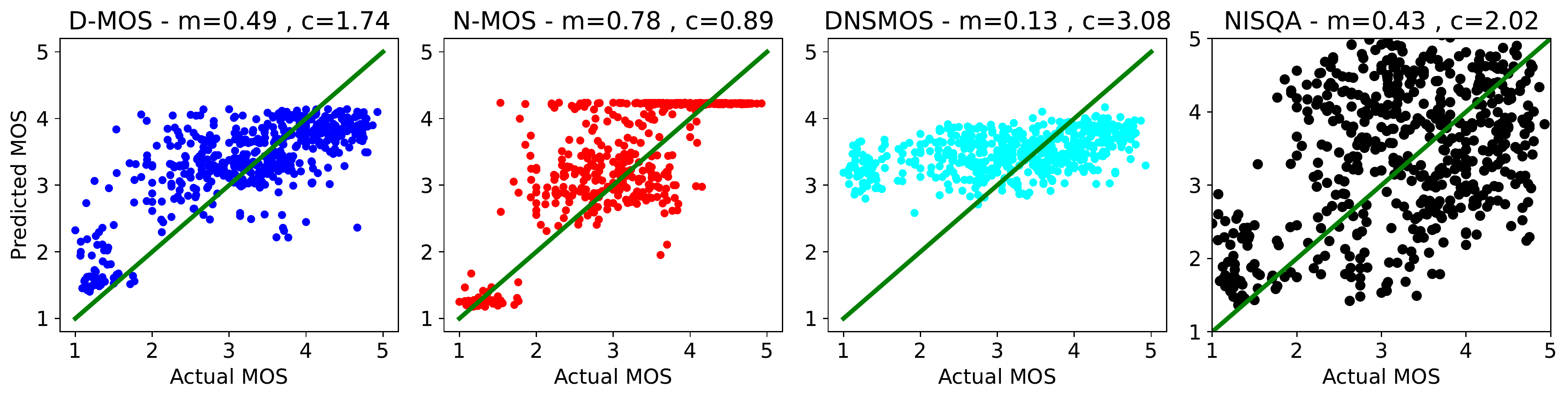} \\

\begin{minipage}{\w}\centering
% \begin{spacing}{0.9}
\centering
{(a) In-Domain Task}
% \end{spacing}
\end{minipage} &

\begin{minipage}{\w}\centering
% \begin{spacing}{0.9}
\centering
{(b) Out-of-Domain Task}
% \end{spacing}
\end{minipage} 

\end{tabular}
\vspace{-2ex}
\caption{\textbf{Scatter plots}: \emph{Top Row}: System level predictions across the \emph{in-domain} and \emph{out-of-domain} tasks of the \VOICEMOS\ challenge dataset (\BVCC). \emph{Bottom Row}: Utterance level predictions across the same dataset and tasks. We show comparisons across \DMOS, \OURS, \DNSMOS, and \NISQA. Green line depicts $y=x$. Each plot title shows the slope (m) and intercept (c) of fitting a linear curve over the points.}
\label{scatter_plot}
\vspace{-1ex}
\end{figure*}

% \begin{figure*}[b!]
% \vspace{-0\baselineskip}
% \centering
% \setlength{\tabcolsep}{4pt}
% \includegraphics[width=0.95\textwidth]{templete/figure/scatter_plots.pdf}
% \vspace{-2ex}
% \caption{\textbf{Utteran}: for \OURS, \DMOS, \DNSMOS, and \NISQA. Mean Square Error (MSE), Spearman (SC), Pearson (PC) correlations are shown. \OURS\, is obtained using $n=100$ NMRs. $\uparrow$ is better.}

% %We use a multi-task learning framework to train a Listening Quality (LQ) head using triplet metric learning strategy, and Spatialization Quality (SQ) using features learnt in predicting source location.}
% % 
% \vspace{-0.1in}
% % \vspace{-0.1in}
% \label{scatter_plot}
% \end{figure*}
\begin{table*}[b!]

\vspace{-0ex}
\centering
\renewcommand{\arraystretch}{1.1}
\resizebox{\textwidth}{!}{
 \begin{tabular}{l l c c c c c c c c c c c c c c c}
 \toprule
  \multirow{2}{*}{\bf Type} & \multirow{2}{*}{\bf Name} & \multicolumn{3}{c}{\bf HiFiGAN~\cite{su2020hifi}} & \multicolumn{3}{c}{\bf VoCo~\cite{jin2017voco}}& \multicolumn{3}{c}{\bf FFTnet~\cite{jin2018fftnet}}& 
 \multicolumn{3}{c}{\bf BWE~\cite{feng2019learning}} & 
 \multicolumn{3}{c}{\bf Dereverb~\cite{su2019perceptually}} \\
 \cmidrule(lr){3-5} \cmidrule(lr){6-8} \cmidrule(lr){9-11} \cmidrule(lr){12-14} \cmidrule(lr){15-17} 
 & &\bf MSE$\downarrow$ &\bf PC$\uparrow$ &\bf SC$\uparrow$ &\bf MSE$\downarrow$ &  \bf PC$\uparrow$ & \bf SC$\uparrow$ &\bf MSE$\downarrow$ & \bf PC$\uparrow$ & \bf SC$\uparrow$ &\bf MSE$\downarrow$ & \bf PC$\uparrow$ & \bf SC$\uparrow$ &\bf MSE$\downarrow$ & \bf PC$\uparrow$ & \bf SC$\uparrow$ \\
 \cmidrule(lr){1-17}

  \multirow{2}{*}{\bf Non-Int.}
  & {\bf DNSMOS} & 0.18 & \bf 0.97 & 0.92 & \bf 0.57 & 0.70 & 0.41 & 0.21 & 0.66 & 0.60 & 1.58 & 0.65 & 0.61 & 2.25 & 0.70 & 0.79 \\
  & {\bf NISQA} & 0.40 & 0.94 & 0.90 & 1.44 & 0.63 & 0.29 & 0.53 & 0.53 & 0.48 & 2.44 & 0.69 & 0.67 & 1.92 & \bf 0.80 & \bf 0.81 \\
 
  \cdashline{1-17}
  
  \multirow{3}{*}{\bf \DMOS}
  & {\bf Scratch} & 0.89 & 0.20 & 0.29 & 1.26 & 0.23 & 0.31 & 0.32 & -0.12 & -0.16 & 1.60 & 0.59 & 0.66 & 2.58 & 0.08 & 0.07  \\
  & {\bf SSL-small} & 0.68  & 0.68 & 0.71 & 0.68 & 0.22 & 0.12 & 0.64 & 0.45 & 0.50 & 0.96 & 0.44 & 0.42 & 1.95 & 0.13 & 0.05 \\
  & {\bf SSL-big} & 0.17 & 0.85 & 0.76 & 1.08 & 0.27 & 0.37 & 1.81 & 0.21 & 0.11 & 2.27 & -0.03 & -0.1 & 2.71 & -0.27 & -0.18\\
  
 \cdashline{1-17}
 
 \multirow{3}{*}{\parbox{1.5cm}{\bf NORESQA-MOS}} 
 & {\bf Scratch} & 0.14 & 0.23 & 0.30 & 0.68 & 0.57 & 0.55 & 0.29 & 0.12 & 0.07 & \bf 0.45 & 0.71 & 0.80 & 1.21 & 0.70 & 0.71 \\
 & {\bf SSL-small} & 0.14 & 0.94 & \bf 0.96 & 0.59 & 0.50 & 0.40 & 0.19 & \bf 0.74 & 0.72 & 0.61 & 0.59 & 0.57 & 0.91 & 0.71 & 0.73  \\
 & {\bf SSL-big} & \bf 0.10 & 0.90 & 0.83 & 0.73 & \bf 0.83 & \bf 0.60 & \bf 0.18 & \bf 0.74 & \bf 0.78 & 1.64 & \bf 0.81 & \bf 0.81 & \bf 0.54 & 0.20 & 0.10\\
 \bottomrule
\end{tabular}
}
\caption{\textbf{System-level-predictions (1)}: for \OURS, \DMOS, \DNSMOS, and \NISQA. Mean Square Error (MSE), Spearman (SC), Pearson (PC) correlations are shown. \OURS\, is obtained using $n=100$ NMRs. $\uparrow$ or $\downarrow$ is better.}
\label{tab:mos_1}
\end{table*}

\noindent {\bf Embedding visualization}:
We visualize how the embedding space looks by projecting the embeddings to a 2D space (Fig~\ref{embeddings_plot}) using dimensionality reduction techniques like PCA~\cite{abdi2010principal} and UMAP~\cite{mcinnes2018umap}. Similar to Manocha \etal~\cite{manocha2021noresqa} we see that the embeddings are more tightly clustered together for higher quality recordings. However, we do not observe two piece-wise linear functions - for low and high quality respectively as in~\cite{manocha2021noresqa}. Instead, we see a continuous projection curve which suggests that the manifold of speech recordings mapped to MOS ratings is smooth without any discontinuities. However, this trend is expected, given the subjective nature of MOS.

% the predictions from the model correlate better with human judgments of quality.

\begin{table*}[t!]

\vspace{-1.0\baselineskip}
\centering
\renewcommand{\arraystretch}{1.1}
\resizebox{\textwidth}{!}{
 \begin{tabular}{l l c c c c c c c c c c c c c c c}
 \toprule
  \multirow{2}{*}{\bf Type} & \multirow{2}{*}{\bf Name} & \multicolumn{3}{c}{\bf TCD\_VOIP~\cite{harte2015tcd}} & \multicolumn{3}{c}{\bf SASSEC~\cite{kastner2019efficient}}& \multicolumn{3}{c}{\bf SiSEC08~\cite{kastner2009evaluating}}& 
 \multicolumn{3}{c}{\bf SiSEC18~\cite{kastner2019efficient}} & 
 \multicolumn{3}{c}{\bf SAOC~\cite{breebaart2008spatial}} \\
 \cmidrule(lr){3-5} \cmidrule(lr){6-8} \cmidrule(lr){9-11} \cmidrule(lr){12-14} \cmidrule(lr){15-17} 
 & &\bf MSE$\downarrow$ &\bf PC$\uparrow$ &\bf SC$\uparrow$ &\bf MSE$\downarrow$ &  \bf PC$\uparrow$ & \bf SC$\uparrow$ &\bf MSE$\downarrow$ & \bf PC$\uparrow$ & \bf SC$\uparrow$ &\bf MSE$\downarrow$ & \bf PC$\uparrow$ & \bf SC$\uparrow$ &\bf MSE$\downarrow$ & \bf PC$\uparrow$ & \bf SC$\uparrow$ \\
 \cmidrule(lr){1-17}
 
\multirow{2}{*}{\bf Non-Int.}
  & {\bf DNSMOS} & 0.71 & \bf 0.96 & \bf 0.96 & 1.37 & 0.74 & 0.81 & 1.24 & 0.54 & 0.56 & 1.06 & 0.22 & 0.23 & 1.45 & 0.64 & 0.64 \\
  & {\bf NISQA} & 0.32 & 0.95 & 0.93 & 0.44 & 0.85 & 0.83 & 1.34 & 0.61 & \bf 0.66 & 1.13 & 0.09 & 0.08 & 1.15 & 0.67 & 0.62 \\
  \cdashline{1-17}
  
  \multirow{3}{*}{\bf \DMOS}
  & {\bf Scratch} & 1.10 & -0.21 & -0.19 & 4.13 & -0.32 & -0.30 & 3.57 & -0.34 & 0.05 & 3.31 & -0.05 & -0.20 & 4.00 & -0.27 & -0.05 \\
  & {\bf SSL-small} & 0.36 & 0.85 & 0.84 & 0.95 & 0.86 & 0.82 & 1.18 & 0.72 & 0.43 & 1.13 & 0.01 & 0.05 & 1.23 & 0.80 & 0.60 \\
  & {\bf SSL-big} & 1.34 & 0.10 & 0.05 & 0.81 & 0.73 & 0.52 & 0.82 & 0.01 & 0.33 & 0.95 & 0.12 & 0.21 & 0.85 & 0.47 & 0.30 \\
  
  \cdashline{1-17}
 
 \multirow{3}{*}{\parbox{1.5cm}{\bf NORESQA-MOS}}
 & {\bf Scratch} & 0.87 & 0.36 & 0.52 & 1.97 & 0.41 & 0.34 & 2.11 & 0.45 & 0.30 & 0.88 & 0.12 & 0.08 & 1.39 & 0.26 & 0.26 \\
  & {\bf SSL-small} & \bf 0.26 & 0.87 & 0.87 & \bf 0.17 & \bf 0.95 & \bf 0.90 & 0.80 & \bf 0.73 & 0.56 & 0.97 & \bf 0.30 & \bf 0.25 & 0.79 & \bf 0.89 & \bf 0.80 \\
  & {\bf SSL-big} & 1.17 & 0.62 & 0.34 & 0.41 & \bf 0.95 & 0.81 & \bf 0.66 & 0.69 & 0.36 & \bf 0.71 & 0.20 & \bf 0.25 & \bf 0.59 & 0.84 & 0.60 \\
 \bottomrule
\end{tabular}
}
\caption{\textbf{System-level-predictions (2)}: for \OURS, \DMOS, \DNSMOS, and \NISQA. Mean Square Error (MSE), Spearman (SC), Pearson (PC) correlations are shown. \OURS\, is obtained using $n=100$ NMRs. $\uparrow$ or $\downarrow$ is better.}
\label{tab:mos_2}
\end{table*}
\begin{table*}[t!]

\vspace{-0.15in}
\centering
\renewcommand{\arraystretch}{1.1}
\resizebox{\textwidth}{!}{
 \begin{tabular}{l l c c c c c c c c c c c c c c c}
 \toprule
  \multirow{2}{*}{\bf Type} & \multirow{2}{*}{\bf Name} & \multicolumn{3}{c}{\bf PEASS\_db~\cite{kastner2019efficient}} & \multicolumn{3}{c}{\bf HiFi2\_Ted~\cite{Su:2021:HSS}}& \multicolumn{3}{c}{\bf HiFi2\_DAPS~\cite{Su:2021:HSS}}& 
 \multicolumn{3}{c}{\bf VoiceMOS\_Main~\cite{cooper2021voices}} & 
 \multicolumn{3}{c}{\bf VoiceMOS\_OOD~\cite{cooper2021voices}} \\
 \cmidrule(lr){3-5} \cmidrule(lr){6-8} \cmidrule(lr){9-11} \cmidrule(lr){12-14} \cmidrule(lr){15-17} 
 & &\bf MSE$\downarrow$ &\bf PC$\uparrow$ &\bf SC$\uparrow$ &\bf MSE$\downarrow$ &  \bf PC$\uparrow$ & \bf SC$\uparrow$ &\bf MSE$\downarrow$ & \bf PC$\uparrow$ & \bf SC$\uparrow$ &\bf MSE$\downarrow$ & \bf PC$\uparrow$ & \bf SC$\uparrow$ &\bf MSE$\downarrow$ & \bf PC$\uparrow$ & \bf SC$\uparrow$ \\
 \cmidrule(lr){1-17}
 
\multirow{2}{*}{\bf Non-Int.}
  & {\bf DNSMOS} & 1.16 & 0.55 & 0.71 & 1.14 & 0.96 & \bf 0.94 & 0.26 & 0.81 & 0.73 & 0.72 & 0.78 & 0.78 & 0.75 & 0.64 & 0.62 \\
  & {\bf NISQA} & \bf 0.26 & 0.36 & 0.25 & 1.19 & 0.94 & \bf 0.94 & 0.91 & 0.76 & 0.72 & 1.06 & 0.67 & 0.71 & 0.70 & 0.60 & 0.54  \\
  \cdashline{1-17}
  
   \multirow{3}{*}{\bf \DMOS}
  & {\bf Scratch} & 3.66 & 0.21 & 0.26 & 0.30 & 0.65 & 0.47 & 0.39 & 0.43 & 0.23 & 0.88 & 0.19 & 0.24 & 0.29 & 0.42 & 0.69 \\
  & {\bf SSL-small} & 0.96 & 0.46 & 0.42 & 0.36 & 0.96 & 0.93 & 0.22 & 0.92 & \bf 0.95 & 0.20 & 0.85 & \bf 0.89 & 0.09 & 0.96 & \bf 0.96 \\
  & {\bf SSL-big} & 0.81 & 0.49 & 0.29 & 0.60 & 0.49 & 0.64 & 0.41 & 0.68 & 0.58 & 0.34 & 0.73 & 0.70 & 0.26 & 0.87 & 0.80   \\
  
  \cdashline{1-17}
 
 \multirow{3}{*}{\parbox{1.5cm}{\bf NORESQA-MOS}}
 & {\bf Scratch} & 0.94 & 0.26 & 0.31 & 0.16 & 0.79 & 0.49 & 0.32 & 0.88 & 0.65 & 0.67 & 0.21 & 0.24 & 0.23 & 0.49 & 0.78 \\
  & {\bf SSL-small} & 0.78 & \bf 0.64 & \bf 0.79& \bf 0.15 & \bf 0.98 & \bf 0.94 & 
  \bf 0.14 & \bf 0.93 & 0.94 & \bf 0.17 & \bf 0.89 & 0.87 & \bf 0.04 & \bf 0.98 & \bf 0.96  \\
  & {\bf SSL-big} & 0.65 & 0.52 & 0.57 & 0.46 & 0.51 & 0.78 & 0.32 & 0.85 & 0.81 & 0.33 & 0.81 & 0.80 & 0.14 & 0.89 & 0.85  \\

 \bottomrule
\end{tabular}
}
\caption{\textbf{System-level-predictions (3)}: for \OURS, \DMOS, \DNSMOS, and \NISQA. Mean Square Error (MSE), Spearman (SC), Pearson (PC) correlations are shown. \OURS\, is obtained using $n=100$ NMRs. $\uparrow$ or $\downarrow$ is better.}
\label{tab:mos_3}
\end{table*}
\begin{table*}[t!]

\vspace{-0.15in}
\centering
\renewcommand{\arraystretch}{1.1}
\resizebox{\textwidth}{!}{
 \begin{tabular}{l l c c c c c c c c c c c c c c c}
 \toprule
  \multirow{2}{*}{\bf Type} & \multirow{2}{*}{\bf Name} & \multicolumn{3}{c}{\bf FFTnet~\cite{su2020hifi}} & \multicolumn{3}{c}{\bf SiSEC18~\cite{jin2017voco}}& \multicolumn{3}{c}{\bf PEASS\_db~\cite{jin2018fftnet}}& 
 \multicolumn{3}{c}{\bf VoiceMOS\_Main~\cite{feng2019learning}} & 
 \multicolumn{3}{c}{\bf VoiceMOS\_OOD~\cite{su2019perceptually}} \\
 \cmidrule(lr){3-5} \cmidrule(lr){6-8} \cmidrule(lr){9-11} \cmidrule(lr){12-14} \cmidrule(lr){15-17} 
 & &\bf MSE $\downarrow$ &\bf PC$\uparrow$ &\bf SC$\uparrow$ &\bf MSE$\downarrow$ &  \bf PC$\uparrow$ & \bf SC$\uparrow$ &\bf MSE$\downarrow$ & \bf PC$\uparrow$ & \bf SC$\uparrow$ &\bf MSE$\downarrow$ & \bf PC$\uparrow$ & \bf SC$\uparrow$ &\bf MSE$\downarrow$ & \bf PC$\uparrow$ & \bf SC$\uparrow$ \\
 \cmidrule(lr){1-17}

 \multirow{2}{*}{\bf Non-Int.}
  & {\bf DNSMOS} & 0.41 & 0.40 & 0.41 & 1.93 & 0.16 & 0.13 & 2.07 & 0.08 & 0.08 & 1.03 & 0.61 & 0.60 & 0.88 & 0.44 & 0.44 \\
  & {\bf NISQA} & 0.92 & 0.37 & 0.25 & 2.56 & 0.10 & 0.05 & 1.38 & 0.12 & 0.11 & 1.45 & 0.62 & 0.63 & 1.12 & 0.43 & 0.33 \\
 
  \cdashline{1-17}
  
  \multirow{3}{*}{\bf \DMOS}
  & {\bf Scratch} & 0.49 & -0.02 & -0.01 & 3.87 & -0.05 & -0.06 & 4.47 & 0.09 & 0.03 & 1.24 & 0.15 & 0.10 & 1.24 & 0.14 & 0.10 \\
  & {\bf SSL-small} & 0.97 & 0.22 & 0.23 & 2.23 & -0.19 & -0.14 & 1.77 & -0.10 & -0.07 & 0.34 & 0.80 & \bf 0.84 & 0.34 & 0.84 & 0.34 \\
  & {\bf SSL-big} & 2.24 & 0.08 & 0.14 & 2.35 & 0.08 & 0.12 & 3.11 & 0.08 & 0.06 & 0.43 & 0.71 & 0.70 & 0.47 & 0.74 & 0.66\\
  
 \cdashline{1-17}
 
 \multirow{3}{*}{\parbox{1.5cm}{\bf NORESQA-MOS}}
 & {\bf Scratch} & \bf 0.31 & 0.01 & 0.02 & 1.64 & 0.14 & 0.13 & 1.76 & 0.14 & 0.12 & \bf 0.88 & 0.18 & 0.13 & 1.02 & 0.28 & 0.19 \\
 & {\bf SSL-small} & 0.72 & 0.36 & 0.36 & \bf 1.12 & \bf 0.20 & \bf 0.20 & \bf 1.16 & \bf 0.18 & \bf 0.13 & 0.31 & \bf 0.83 & 0.83 & \bf 0.29 & \bf 0.85 & \bf 0.81 \\
 & {\bf SSL-big} & 0.50 & \bf 0.51 & \bf 0.48 & 2.09 & 0.16 & 0.12 & 2.73 & 0.10 & 0.06 & 0.47 & 0.76 & 0.73 & 0.43 & 0.76 & 0.73 \\
 \bottomrule
\end{tabular}
}
\caption{\textbf{Utterance-level-predictions}: for \OURS, \DMOS, \DNSMOS, and \NISQA. Mean Square Error (MSE), Spearman (SC), Pearson (PC) correlations are shown. \OURS\, is obtained using $n=100$ NMRs. $\uparrow$ or $\downarrow$ is better.}
\label{ablations_1}
\vspace{-5.9ex}
\end{table*}

\vspace{-1\baselineskip}
\subsection{Subjective evaluations}
\label{sub_eval}
We evaluate MOS prediction through an exhaustive set of 16 different datasets. These datasets come from a variety of speech applications including speech synthesis (VoCo~\cite{jin2017voco}, FFTnet~\cite{jin2018fftnet}), speech enhancement (Dereverberation~\cite{su2019perceptually}, HiFi-GAN~\cite{su2020hifi}, HiFi-GAN2~\cite{Su:2021:HSS}), audio source separation (SASSEC~\cite{kastner2019efficient}, SiSEC08~\cite{kastner2009evaluating}, SiSEC18~\cite{kastner2019efficient}, SAOC~\cite{breebaart2008spatial}), telephony degradations (TCD\_VOIP~\cite{harte2015tcd}), bandwidth extension (BWE~\cite{feng2019learning}), and Voice Conversion and TTS (\BVCC~\cite{voicemos}). 
%
%\AKnote{For each of these, put example datasets in brackets. E.g. bandwidth expansion (BWE [CITATION])}. 
%
For more information on the datasets, please refer to Manocha \etal~\cite{Manocha:2022:SNS}. Our goal is to establish the generalization capabilities of all methods by evaluating on these diverse datasets. 
%Note that our models are trained on \VOICEMOS\ dataset where the speech recordings are limited to just two domains TTS and Voice-Conversion.

Similar to prior works, we measure performance through Mean Square Errors (MSE), Pearson Correlation Coefficient (PC), and Spearman’s Rank Order Correlation (SC) of our predicted MOS with the MOS ratings from each dataset.
The NMRs for \OURS\ are selected randomly from DAPS dataset~\cite{mysore2014can}. For \OURS, all experiments are repeated 10 times and averaged results with standard deviations are reported. We report both \emph{system level} (averaged over ratings per system), as well as the \emph{utterance level} predictions.

% \AKnote {In a line describe the system and utterance level ratings here.}

\noindent {\bf Scatter plots}
Fig~\ref{scatter_plot} shows the performance of various metrics on a common dataset (\BVCC) at a system level, and at an utterance level on in-domain and out-of-domain tasks.
%We show the performance of various metrics on a common dataset (\BVCC) at a system level, as well as at an utterance level to show how well different metrics correlate with subjective ratings. Refer to Fig~\ref{scatter_plot}.

%\AKnote{Are you showing both system and utterance level scatter plots ? If not dont mention anything here. Also there are 4 sets of plots here. Which datasets are these ?}

We see that \OURS\ correlates better than existing baselines including \DMOS. Looking at system level ratings, our approach has a smaller variance spread as compared to baseline approaches. Next, looking at utterance level ratings, we see that baseline approaches have either higher variance (\NISQA\ and \DMOS) or high bias (\DNSMOS). This broadly suggests the usefulness of our approach over existing approaches.
%\AKnote{Give a proper caption to the figure. What are the 4 sets of plots ? which datasets ? system/utterance?}

\noindent {\bf System level MOS predictions}
%\AKnote{Predictions instead of correlations ??, we have MSE too in the tables}}
Results are displayed in Tables~\ref{tab:mos_1},~\ref{tab:mos_2} and~\ref{tab:mos_3}.
%
% \begin{enumerate}[leftmargin=0.33cm]
%     \setlength{\itemsep}{0pt}
%     \setlength{\parskip}{0pt}
%     \setlength{\parsep}{0pt}
We note a few key observations from the these Tables. First, we note that \OURS\ performs better than \DMOS\ across \emph{all} three model classes. 
%\AKnote{Make it consistent. Whatever you call it Abs. MOS or D-MOS or whatever, be consistent everywhere - tables, captions, text. There are mismatches all over the place}. 
We attribute this to our NMR strategy that encourages learning content agnostic quality features. For \eg, specifically for \textit{Dereverb} - \DMOS\ models fare worse than \OURS\ because they fail to give reliable estimates, esp. under unseen, reverberant environments. In contrast, \OURS\ performs better since it was trained to be content agnostic to learn quality features.
Secondly, we also observe that generalization across unseen datasets generally increase with larger pretrained SSL models (\eg\ HiFi-GAN, SASSEC, SiSEC08, SiSEC18, SAOC \etc). However, in a few cases, the performance drops as larger pretrained models are used, esp. for \DMOS. 
%This may be due to the instability caused due to vanishing gradients in large models~\cite{mosbach2020stability} during finetuning. %% don't think we can attribute it to vanishing gradient.
However, our \OURS\ approach produces more consistent ratings across model capacities.
Third, we note that \OURS\ with the \textit{SSL-Small} model generalizes better than \DMOS\ learning with \textit{SSL-Big} as base modules. %\AKnote{SSL-Big/Small also consistent everywhere, in text (section 2 and here) as well as all tables/results}. 
This shows the usefulness of our approach in training efficient models (\ie\ with 1/4 the number of trainable parameters) that generalize well, and are faster to train and infer. 
Fourth, \OURS\ approach scores higher than baseline approaches like \DNSMOS\ and \NISQA\ in terms of lower errors (MSE) and higher correlations, especially on challenging datasets like \emph{BWE} which have subtle differences. 
The standard deviations for all datasets across Tables~\ref{tab:mos_1},~\ref{tab:mos_2}, and~\ref{tab:mos_3} are consistently small ($\sim$0.02 rating) suggesting invariance to a particular NMR set.
Moreover, \OURS\ uses a fraction of the labeled examples for training compared to \DNSMOS\ or \NISQA\ and therefore is more effective for sparse labeled tasks.
Finally, we note that our NMRs based MOS estimation approach improves performance across all model classes, whether training from scratch, or starting from a pretrained model across various model capacities. It shows that our approach is a generic way to improve MOS estimation and can be used to improve robustness of \emph{any} model.

\noindent {\bf Utterance level MOS predictions} 
%\AKnote{again predictions ??}}
We report results on a subset of datasets from the previous section due to space limitations.

Results are shown in Table~\ref{ablations_1}. We see that \OURS\ scores consistent correlations, and lowest errors amongst different datasets considered. Moreover, the standard deviations for datasets across Table~\ref{ablations_1} are small ($\sim$0.15 rating), and should further decrease as more NMRs are introduced. This suggests the usefulness of our approach to reducing variance in the ratings further.
Utterance level MOS predictions have been identified as challenging for existing models~\cite{cooper2021generalization}. Our \OURS\ approach can produce more consistent ratings and improves performances almost across all datasets.

\vspace{-0.01in}
\section{Conclusions and future work}
In this paper, we presented \OURS\ - a novel approach for MOS estimation of speech signals which uses non-matching references. It is motivated by human’s ability to assess quality independent of the speech content. We show that our method generalizes well to out-of-domain datasets
and outperforms prior works trained on much larger datasets. Moreover, it provides good generalization with smaller models, making it more suitable for real-world uses. In the future, we would like to include more attributes including noisiness and coloration.

\clearpage
\pagebreak
\bibliographystyle{IEEEtran}
\bibliography{mybib}

\end{document}